\documentclass[pra,aps,epsf,nofootinbib,
notitlepage,showpacs]{revtex4-2}

\usepackage{color}
\usepackage{float}
\usepackage{graphicx}
\usepackage{subfigure}
\usepackage[normalem]{ulem} 
\usepackage{amsmath}
\usepackage{amsthm}
\usepackage{mathtools}

\theoremstyle{definition}

\usepackage{bm}
\usepackage{hyperref}
\hypersetup{
     colorlinks   = true,
     citecolor    = blue
}

\newcommand{\beq}{\begin{equation}}
\newcommand{\eeq}{\end{equation}}
\def\be*{\begin{equation*}}
\def\ee*{\end{equation*}}
\def\bea{\begin{eqnarray}}
\def\eea{\end{eqnarray}}
\usepackage[normalem]{ulem}

\newcommand{\bra}[1]{{\left\langle #1 \right|}}
\newcommand{\ket}[1]{{\left| #1 \right\rangle}}

\theoremstyle{definition}

\usepackage{graphicx} 
\graphicspath{ {rrImages/} } 
\usepackage{tikz} 

\usepackage[]{xcolor}

\usetikzlibrary{matrix,positioning}
\usepackage{amsfonts}
\usepackage{amssymb}
\newcommand{\ra}[1]{{\color{black}{#1}}} 
\definecolor{myblavender}{rgb}{0.961, 0.456, 1.0}

\newcommand{\cmmnt}[1]{}
\begin{document}


\title{Solving MAXCUT with Quantum Imaginary Time Evolution}
	
\author{Rizwanul Alam}
\email{ralam4@tennessee.edu}
\affiliation{
	Department of Physics and Astronomy, University of Tennessee at Knoxville\\Knoxville, Tennessee 37996-1200 USA}
	
\author{George Siopsis}
\email{siopsis@tennessee.edu}
\affiliation{
	Department of Physics and Astronomy, University of Tennessee at Knoxville\\Knoxville, Tennessee 37996-1200 USA}
	
\author{Rebekah Herrman}
\email{rherrma2@tennessee.edu}
\affiliation{
	Department of Industrial and Systems Engineering, University of Tennessee at Knoxville\\Knoxville, TN  37996 USA}

\author{James Ostrowski}
\email{jostrows@tennessee.edu}
\affiliation{
	Department of Industrial and Systems Engineering, University of Tennessee at Knoxville\\Knoxville, TN  37996 USA}

\author{Phillip C. Lotshaw}
\email{lotshawpc@ornl.gov}
\affiliation{
	Quantum Computing Institute\\ Oak Ridge National Laboratory\\ Oak Ridge, TN 37830 USA}
	
\author{Travis S.~Humble}
\email{humblets@ornl.gov}
\affiliation{
	Quantum Computing Institute\\ Oak Ridge National Laboratory\\ Oak Ridge, TN 37830 USA}

\date{December 2021}

\begin{abstract}
We introduce a method to solve the MaxCut problem efficiently based on quantum imaginary time evolution (QITE). We employ a linear \textit{Ansatz} for unitary updates and an initial state involving no entanglement, as well as an imaginary-time-dependent Hamiltonian interpolating between a given graph and a subgraph with two edges excised. We apply the method to thousands of randomly selected graphs with up to fifty vertices.  We show that our algorithm exhibits a 93\% and above performance converging to the maximum solution of the MaxCut problem for all considered graphs. Our results compare favorably with the performance of classical algorithms, such as the greedy and Goemans-Williamson algorithms. We also discuss the overlap of the final state of the QITE algorithm with the ground state as a performance metric, which is a quantum feature not shared by other classical algorithms. This metric can be improved by introducing higher-order \textit{Ans\"atze} and entangled initial states.   \end{abstract}

\maketitle
\section{Introduction}

Since fault-tolerant universal quantum computers have yet to be developed, considerable effort is focused on demonstrating quantum advantage with currently available noisy intermediate-scale quantum (NISQ) devices, including superconducting \cite{arute2019quantum} and photonic \cite{Zhongeabe8770} quantum computers. An area of practical importance in these explorations is finding approximate solutions to combinatorial optimization problems, such as MaxCut. Finding exact solutions to MaxCut is classically hard, but near optimal solutions can be found classically \cite{festa2002randomized, 10.1007/3-540-63774-5_137,articlegreedy,10.5555/1347082.1347102,7133122,goemans1994879}. Quantum algorithms promise a speedup over classical ones. However, it is a challenge to demonstrate their advantage with NISQ devices. 

A widely studied quantum algorithm for combinatorial optimization problems which is suitable for NISQ hardware is the quantum approximate optimization algorithm (QAOA) \cite{farhi2014quantum}. It has been discussed both theoretically and experimentally \cite{wang2018quantum,Hadfield2018dissertation,zhou2020quantum,MaxCutRequiresHundredsQubits,medvidovic2021classical,brandao2018concentration,Love2020Bounds,Shaydulin2020CaseStudy,crooks2018performance,Shaydulin2020Symmetries,herrman2021lower,ReachabilityDeficit,Szegedy2020GraphQAOA,Pagano2020TrappedIonQAOA}. Variants of QAOA have also been explored \cite{wang2020xy,zhu2020adaptqaoa,Jiang2017GroverQAOA,Eidenbenz2020GroverMixers,Bartschi2020MaxkCover,LiLi2020Gibbs,Gupta2020WarmStart,Herrman2022-wh}. Motivated by adiabatic evolution, QAOA uses a string of unitary evolution operators alternating between two Hamiltonian functions with time parameters that are optimized classically in order to maximize the cost function (equivalently, minimize the energy of the corresponding Hamiltonian). One starts with a state which is not entangled and the desired final state (ground state) need not be entangled.  However, the quantum circuit introduces entanglement which is expected to provide quantum advantage in the calculation of the minimum energy eigenvalue. In practice, it is challenging to establish quantum advantage, in the absence of a theoretical argument, because only shallow quantum circuits can be implemented on NISQ hardware without overwhelming quantum errors.

A popular class of classical algorithms applied to MaxCut are the greedy algorithms that rely on greedy vertex labeling or an edge contraction strategy \cite{10.1007/3-540-63774-5_137,articlegreedy,10.5555/1347082.1347102,7133122}. They exhibit a 50\% performance for the worst-case graphs. The best worst-case performance is provided by the Goemans-Williamson (classical) algorithm \cite{goemans1994879} at 88\% approximation ratio for MaxCut.
It is instructive to look at the performance of QAOA compared to other classical algorithms. It has been shown that \textit{quantum-inspired optimization} methods, such as the local tensor method, outperform single-step QAOA on triangle free graphs \cite{hastings2019classical, bapat2021approx}. There are other studies \cite{Marwaha2021localclassicalmax} showing that local classical MaxCut algorithms outperform $p=2$ QAOA on regular graphs of $\mbox{girth} \geq 5$. In a study of more generalized problems, e.g., approximately solving instances of Max \textit{k}XOR, calculations have been performed to obtain  numerical upper and lower bounds on local classical and quantum algorithms for triangle-free instances \cite{Marwaha2022Bounds}. 

Here we discuss a different method to solve combinatorial optimization problems, focusing on MaxCut, based on quantum imaginary time evolution (QITE). The QITE algorithm has been widely used to find the ground-state energy of many-particle systems \cite{McArdle2019,beach2019making}. Since evolution in imaginary time effectively cools the system down to zero temperature \cite{love2020cooling}, the ground state can be prepared with QITE exactly without any variational optimization. However, in practice, due to limited computational resources, approximations must be made calling for an approach involving variational calculus. An approach to QITE for the computation of the energy spectrum of a given Hamiltonian was outlined in  \cite{motta2020determining}. It had the advantage compared to a variational quantum eigensolver (VQE) of not using ancilla qubits. The method was applied to the quantum computation of chemical energy levels on NISQ hardware \cite{gomes2020efficient, yeter2020practical, yeter2021benchmarking,2020arXiv201108137B}, and the simulation of open quantum systems \cite{kamakari2021digital}. The impact of noise on QITE in NISQ hardware was addressed in \cite{ville2021leveraging} using error mitigation and randomized compiling. Error mitigation was also addressed with a different method based on deep reinforcement learning \cite{cao2021quantum}. A reduction of the depth of quantum circuits for QITE using a nonlocal approximation was discussed in \cite{nishi2021}. Real and imaginary time evolution with compressed quantum circuits on NISQ hardware were performed in \cite{PRXQuantum.2.010342}.



For the MaxCut problem, QITE can be formulated to introduce entanglement similar to QAOA. Moreover, entanglement may also be present in the initial state, which is arbitrary when the QAOA evolution operators are adjusted appropriately \cite{hadfield2019quantum}, as long as it has finite overlap with the ground state. Following \cite{motta2020determining}, we implement QITE in a string of small steps each involving a unitary update. Similar to QAOA, the unitary operator in each QITE step involves variational parameters, but unlike QAOA, these parameters are fixed by algebraic equations. We chose an initial state and unitary updates that contained no entanglement. This gives a classical baseline for QITE performance, which can be compared to entangling \textit{Ans\"atze} in future work to assess the role of entanglement. We applied the QITE method with these choices to the MaxCut problem on graphs with up to fifty vertices. Remarkably, within  eleven QITE steps on average, we obtained solutions which were on average at  89\% or better of the optimal solution. As the number of vertices increased from  eight to fifty, the performance of the algorithm remained high dropping from  99\% to 89\%. Regarding efficiency, each QITE step involves the solution of algebraic equations with number of manipulations $\mathcal{O} (|V|^2)$. The number of QITE steps needed also appears to depend polynomially on the number of vertices $|V|$, although further analysis of higher-order graphs is needed to better determine this dependence. These results indicate that our linear QITE method is efficient and quantum advantage due to entanglement is likely to be found only at larger graphs requiring deep quantum circuits which cannot currently be handled by NISQ hardware.


Moreover, a slight modification of our method which also introduced no entanglement, led to  above 93\% performance for all graphs with up to fifty vertices that we studied. The modification involved an imaginary-time-dependent Hamiltonian interpolating between the given graph and a subgraph with two edges excised within a few QITE steps. Identifying the two edges resulting in evolution leading to the ground state introduces a polynomial overhead in the algorithm. Further work with larger graphs is required to identify the point of failure of this modified QITE method.

Our discussion is organized as follows. In Section \ref{sec:2}, we introduce the QITE method with our linear \textit{Ansatz} that involves no entanglement applied to the MaxCut problem. In Section \ref{simulations}, we present our results on thousands of randomly selected graphs with up to fifty vertices showing that our method always works on all the graphs we studied.
Finally in Section \ref{conclusion}, we offer concluding remarks.



\section{QITE for MAXCUT}\label{sec:2}

In this Section, we introduce our QITE method applied to the MaxCut problem. We employ a linear \emph{Ansatz} for unitary updates which introduces no entanglement. Numerical results are presented in the next Section, where we apply the method to thousands of randomly selected graphs with up to fifty vertices. Remarkably, despite performing approximations at each QITE step, the method always converged to the optimal solution of the MaxCut problem for all the graphs we examined. 

Given a graph $G = (V,E)$ consisting of a set of vertices $V$ and edges $E \subseteq V\times V$ joining the vertices in $V$, the MaxCut problem on $G$ is the combinatorial optimization problem of partitioning $V$ into two disjoint sets such that the number of edges with endpoints in each set, $C$, is maximized ($C = C_{\text{max}}$). It can be formulated as a Hamiltonian ground-state problem by associating a qubit with every vertex in $V$ and defining the Hamiltonian
\begin{equation} \mathcal{H} = \sum_{(ij) \in E} Z_i Z_j \end{equation}
where $Z_i$ is a Pauli $Z$-matrix acting on the qubit at the $i$th vertex. Here, the eigenstates of $\mathcal{H}$ are computational basis states $\vert z \rangle = \bigotimes_{j=1}^{|V|} \vert z_j\rangle$, where $z_j  \in \{ 0, 1 \}$, and $|z_j\rangle$ is the state of the qubit at the $j$th vertex. The solution $C_\text{max}$ to the MaxCut problem is related to the ground-state energy $\mathcal{E}_{0}$ of $\mathcal{H}$ by
\beq C_{\text{max}} = \frac{|E| - \mathcal{E}_0}{2} \eeq
All eigenvalues of the Hamiltonian correspond to solutions of the MaxCut problem which are not necessarily optimal,
\beq\label{eq:3} C_k = \frac{|E| - \mathcal{E}_k}{2} \ , \ \ k= 0,1,\dots, 2^{|V|} -1 \eeq
Evidently, $C_k / C_{\text{max}} \le 1$.


To find the ground-state energy, the QITE algorithm relies on the fact that any state with non-vanishing overlap with the ground state eventually reduces to the ground state if it is evolved in imaginary time. In other words, the state
\beq |\Omega\rangle = \lim_{\beta\to\infty} \ket{\Psi (\beta)} \ , \ \ \ket{\Psi(\tau)} \equiv  \frac{e^{-\tau \mathcal{H}}|\Psi\rangle}{\|  e^{-\tau \mathcal{H}} |\Psi\rangle \|} \label{eq:QITE}\eeq
is the ground state for any state $|\Psi\rangle$, as long as $\langle \Omega |\Psi\rangle \ne 0$,
\beq \mathcal{H} |\Omega\rangle = \mathcal{E}_0 |\Omega\rangle \eeq The imaginary time parameter $\beta$ can also be thought of as the inverse temperature ($\beta = \frac{1}{k_B T}$, where $T$ is temperature and $k_B$ is the Boltzmann constant), in which case Eq.\ \eqref{eq:QITE} states that the system settles to the ground state at zero temperature.

To implement \eqref{eq:QITE}, 
we perform evolution in small imaginary time intervals $\tau$. Starting with $|\Psi[0]\rangle$, suppose that after $s-1$ steps we arrive at the state $|\Psi[s-1]\rangle$.  At the next ($s$th) step, we wish to construct the state
\beq\label{eq:9} |\Psi'(\tau)\rangle = \frac{e^{-\tau \mathcal{H} } |\Psi[s-1]\rangle}{\| e^{-\tau \mathcal{H} } |\Psi[s-1]\rangle \|} \eeq
The state $\ket{\Psi'(\tau)}$ has lower energy than $\ket{\Psi[s-1]}$ for sufficiently small $\tau$. To see this, calculate the derivative of the average energy with respect to $\tau$. We obtain
\beq \left. \frac{d}{d\tau} \bra{\Psi' (\tau)} \mathcal{H}\ket{\Psi'(\tau)} \right|_{\tau = 0} = - 2(\Delta E [s-1])^2 \eeq
where $\Delta E[s-1]$ is the uncertainty in energy ($\Delta E = \sqrt{\langle{\mathcal{H}^2}\rangle - \langle{\mathcal{H}}\rangle^2}$) in the state $\ket{\Psi [s-1]}$. It follows that the derivative is negative and the energy is a decreasing function at $\tau = 0$. This step does not decrease the energy if the uncertainty vanishes, $\Delta E [s-1] =0$. This is the case when the state $\ket{\Psi [s-1]}$ is an eigenstate of the Hamiltonian. If the eigenstate is the ground state, then no further steps are needed. However, it can be an excited state, and then the algorithm fails to reach the ground state. It would be desirable to understand how the structure of graphs influences the final state and how one can overcome convergence to an excited energy state.

As we will see, in order to avoid reaching an excited state, it is advantageous to interpolate between a Hamiltonian that corresponds to a subgraph of $G$ and $\mathcal{H}$. We therefore define
\beq\label{eq:Hbeta} \mathcal{H}[s] \equiv \sum_{(ij) \in E} h_{ij} [s] Z_i Z_j \eeq
where all $h_{ij} [s] \to 1$ for large enough $s$ (say, $h_{ij} [s] =1$, for $s\ge s_0$), so that $\mathcal{H}[s] \to \mathcal{H}$.
To select a given subgraph of $G$ as starting point, the coefficients that do not correspond to an edge in the subgraph are set to vanish initially as described further in the next Section.

The state $\ket{\Psi'(\tau)}$ is approximated by a unitary acting on $\ket{\Psi[s-1]}$, $e^{-i\tau A[s]}$, where $A[s]$ is a Hermitian operator. Then after $s$ steps, we arrive at the state 
\beq |\Psi [s]\rangle = e^{-i\tau A[s]} |\Psi[s-1]\rangle \label{eq:approx}\eeq
This is done by minimizing the distance $\delta = \| |\Psi[s]\rangle - |\Psi' (\tau)\rangle \|$ between the approximately evolved state \eqref{eq:approx} and the desired state, \eqref{eq:9},
where 
we used the definition $\| |\Phi \rangle \| = \sqrt{\langle \Phi |\Phi \rangle}$ for the norm of a state $|\Phi\rangle$. At first order in the small imaginary time parameter $\tau$, minimizing $\delta$ leads to a linear system of algebraic equations that can be used to determine the free parameters in $\mathcal{A} [s]$. We may expand $\mathcal{A} [s] = \mathcal{A}_1 [s] + \mathcal{A}_2 [s] + \dots$, where $\mathcal{A}_k [s]$ is a linear combination of products of $k$ Pauli matrices ($k=1,2,\dots$). Including higher values of $k$ brings the distance $\delta$ closer to zero at the expense of increasing the complexity of the calculation. To perform a systematic study, we start by including terms with $k=1$ only, and leave the inclusion of higher-order terms in $\mathcal{A} [s]$ to future work.

Thus, to determine this unitary update, we adopt the \emph{linear Ansatz}
\beq A[s] = \sum_{j\in V} a_j[s] Y_j \label{eq:8}\eeq
where $Y_j$ is the $Y$-Pauli matrix acting on the qubit at the $j$th vertex.
It is straightforward to see \cite{motta2020determining} that the distance $\delta$ is minimized for coefficients $a_j[s]$ obeying the linear system of equations 
\beq\label{eq:10} \bm{S}\cdot \bm{a} = \bm{b} \ , \ \ S_{ij}[s] = \langle Y_i Y_j \rangle \ , \ \ b_j[s] = -\frac{i}{2} \langle [ \mathcal{H} , Y_j] \rangle \eeq
where all expectation values are evaluated with respect to the state $|\Psi[s-1]\rangle$ obtained in the previous step. Notice that the commutator in $\bm{b}$ can be written as
\beq [ \mathcal{H} , Y_j] = -2i H_j \mathcal{H}_{G_j} H_j \eeq
where we used $HZH = X$ with $H_j$ being the Hadamard matrix $H$ acting on the qubit at the $j$th vertex, and $G_j$ the subgraph of $G$ consisting of the vertex $j$ and its adjacent vertices in $G$ with Hamiltonian
\beq\label{eq:H} \mathcal{H}_{G_j}  = \sum_{(ij) \in E(G_j)} h_{ij}  Z_i Z_j \eeq
Since $\mathcal{H}_{G_j}$ is diagonal in the computational basis, $b_j [s]$ can be computed by engineering the state $H_j |\Psi[s-1]\rangle$, measuring each qubit, and using
\beq b_j[s] = - \langle \Psi[s-1] |H_j \mathcal{H}_{G_j} H_j |\Psi[s-1] \rangle \eeq
Similarly, the matrix elements of $\bm{S}$ can be expressed in terms of expectation values involving the two-qubit matrix $Z_iZ_j$ which is diagonal in the computational basis,
\beq S_{ij}[s] = \langle \Psi[s-1]| e^{i\frac{\pi}{4}(X_i + X_j)} Z_iZ_j e^{-i\frac{\pi}{4}(X_i+X_j)} |\Psi[s-1]\rangle \eeq
where we used $e^{i\frac{\pi}{4}X} Z e^{-i\frac{\pi}{4}X} = Y$, and can be obtained by engineering the state $e^{-i\frac{\pi}{4}(X_i+X_j)} |\Psi [s-1]\rangle$ and measuring all qubits.

It should be noted that the unitary updates \eqref{eq:approx} with the linear \emph{Ansatz} \eqref{eq:8} do not introduce entanglement. If one starts with a separable initial state, the state at each step in the QITE algorithm will be separable. A further simplification occurs if the initial state is chosen to be the tensor product of eigenstates of $X$ and $Z$,
\beq |\Psi[0]\rangle = \bigotimes_{j=1}^{|V|} |s_j\rangle \eeq 
where $|s_j\rangle \in \{ |0\rangle, |1\rangle, |+\rangle, |-\rangle \}$, and $\ket{\pm} = \frac{1}{\sqrt{2}} (\ket{0} \pm \ket{1})$. In this case, the matrix $\bm{S}$ is the identity, because $\langle s_j | Y_j |s_j\rangle =0$. It follows that $\bm{a} = \bm{b}$. It is easy to see that single-qubit states should not all be eigenstates of $X$, because they yield $\bm{a} = \bm{0}$, due to $\langle \pm |Z|\pm\rangle = 0$ in \eqref{eq:10}. 
Let us change the state of the qubit at position $j$ to $|0\rangle$. Then we obtain non-vanishing coefficients $a_i[s]$ for the qubits that are adjacent to $j$, i.e., $a_i[s] \ne 0$ for $(ij) \in E$. At the next step in the QITE algorithm, we obtain non-vanishing coefficients for all qubits at distance up to 2 from the one at position $j$. For a connected graph, it takes less than $|V|$ steps to obtain non-vanishing coefficients for all qubits.

After $s$ steps in the QITE algorithm, we arrive at the state
\beq |\Psi[s]\rangle = e^{-i\tau \mathcal{A}[s]} |\Psi[0]\rangle \ , \ \ \mathcal{A}[s] = \sum_{s'=1}^s A[s'] \eeq
%
%
In our calculations, we chose an initial state in which all qubits were set to $|+\rangle$, except one which was set to $|0\rangle$. We chose the latter qubit to be at one of the highest-degree vertices. This is an arbitrary choice, but appears to be more efficient in some cases. Thus,
\beq\label{eq:4} |\Psi[0]\rangle = H_j |\Psi_0\rangle \ , \ \ |\Psi_0\rangle = |+\rangle^{\otimes |V|} \eeq
where $H_j$ is the Hadamard matrix at the position of the chosen qubit, and $|\Psi_0\rangle$ is the standard initial choice in QAOA.

At the $s$th QITE step, we used the unitary update $e^{-i\tau \mathcal{A} [s]}$ and optimized the choice of the small imaginary-time parameter $\tau$ by minimizing the average energy $\langle \mathcal{H} \rangle$ of the state at the $s$th step. Note that this involves computing all the intermediary states $\vert \Psi[s']\rangle$ to evaluate the coefficients $a_j[s']$ in \eqref{eq:10}, for all $s'<s$. 

Regarding efficiency, we note that at each QITE step, we need to solve a system of algebraic equations which requires a number of manipulations which is polynomial in the number of vertices $|V|$ of the given graph.  Moreover, as we show in Section \ref{simulations}, with just 5 QITE steps it is possible to achieve more than $88\%$ of $C/C_{\text{max}}$. With an average of 10 steps, it goes over $93\%$ for $|V| \leq 50$. Performance can be further improved with more steps and by making the Hamiltonian imaginary-time dependent, as discussed in Section \ref{simulations}. On the metric of ground-state overlap, we find that $75\%$ or more graphs in the random sample we considered have non-zero overlap with the ground state. This can also be further improved by increasing the number of steps and a thorough study of imaginary-time dependent Hamiltonian.  Although more work is needed with higher-order graphs, these results point to an efficient method of solving the MaxCut problem for graphs  with up to at least 50 vertices, that does not require unitary updates with a higher-order \textit{Ansatz}.

\section{Results}\label{simulations}

\begin{figure}[ht!]
    \centering
        \subfigure[]{\includegraphics[scale=.3]{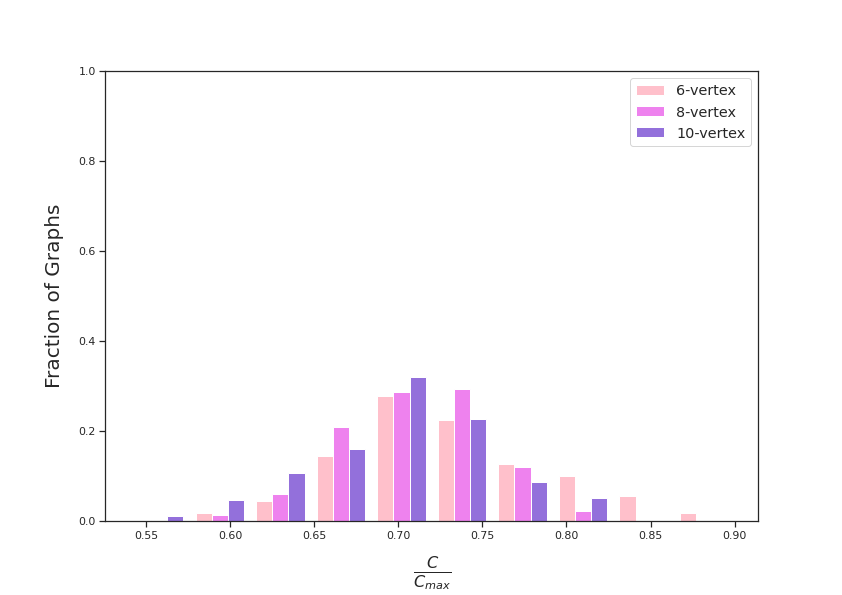}}
        \subfigure[]{\includegraphics[scale=.3]{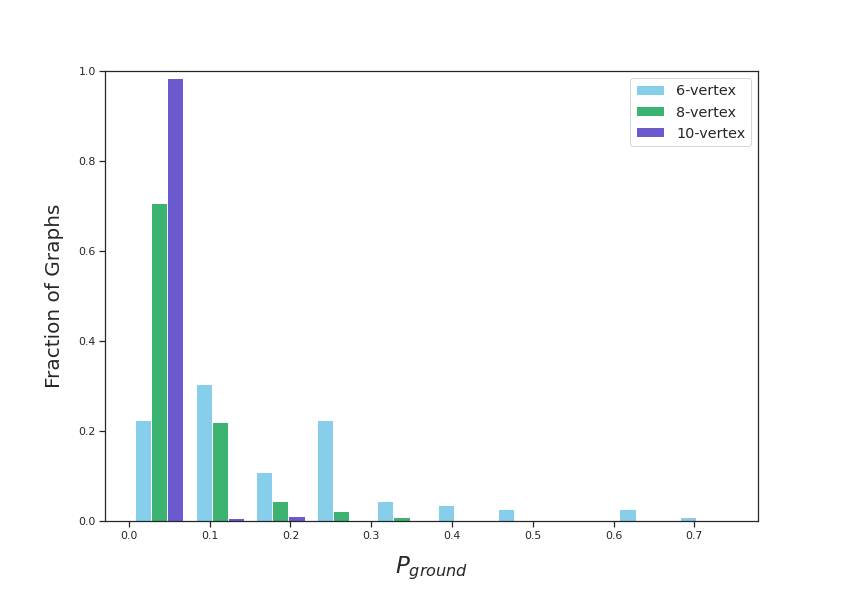}}
    \caption{Histograms of (a) $C/C_{\text{max}}$ and (b) probability of overlap with ground state $P_{\text{ground}}$ after $s=1$ QITE step with fixed Hamiltonian for all connected six- and eight-vertex graphs, and 200 randomly chosen connected ten-vertex graphs.  The average values of $C/C_{\text{max}}$ are 0.73, 0.71, and 0.70, for six-, eight-, and ten-vertex graphs, respectively. The corresponding average values of $P_{\text{ground}}$ are 0.2, 0.06, and 0.02. 
    }
    \label{fig:C1P1}
\end{figure}

In this Section, we present our results of applying the QITE algorithm with a linear \emph{Ansatz} to graphs with up to 
fifty vertices.

We applied our algorithm to all graphs with up to eight vertices since there is a computationally manageable number of such graphs (e.g., there are only 11,117 distinct graphs with eight vertices). We employed two metrics to assess the performance of our algorithm. One is the average value of $C/C_{\text{max}}$ which is called the approximation ratio and is a metric shared by both classical and quantum algorithms. The other is the overlap of the final state with the ground state that leads to the probability $P_{\text{ground}}$ that a measurement yields the ground state of the Hamiltonian, and therefore the optimal solution $C_{\text{max}}$ to the MaxCut problem.

Figure \ref{fig:C1P1} shows the results of applying one step of the QITE algorithm with a linear \emph{Ansatz} to all connected six- and eight-vertex graphs, and 200 randomly selected ten-vertex graphs. We used a fixed Hamiltonian, setting all coefficients $h_{ij} =1$ in Eq.\ \eqref{eq:H}. We obtained average values of $C/C_{\text{max}}$ 73\%, 71\%, and 70\%, for six-, eight-, and ten-vertex graphs, respectively. For the probability $P_{\text{ground}}$   we obtained corresponding average values of 20\%, 6\%, and 2\%. In terms of the latter metric, the performance drops significantly as the number of vertices in the graphs increases.

\begin{figure}[ht!]
    \centering
        \subfigure[]{\includegraphics[scale=.3]{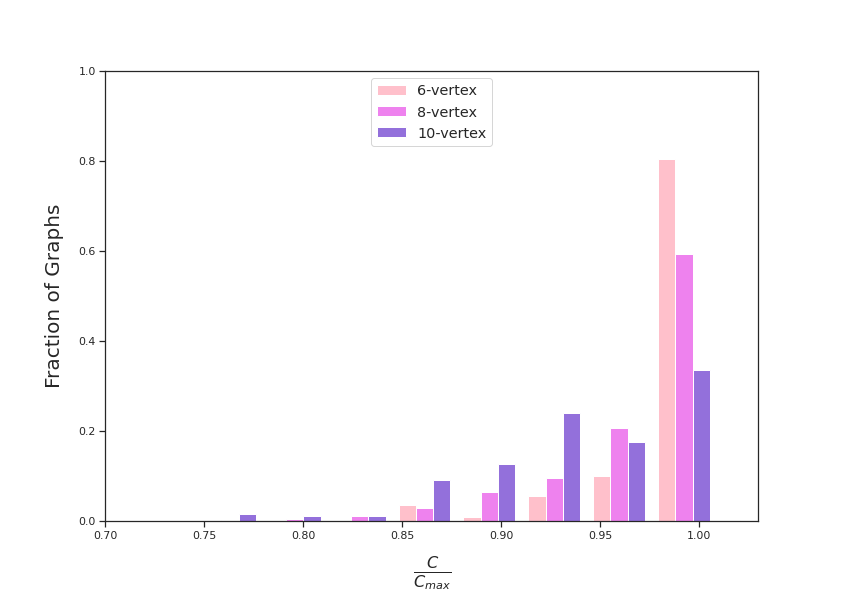}}
        \subfigure[]{\includegraphics[scale=.3]{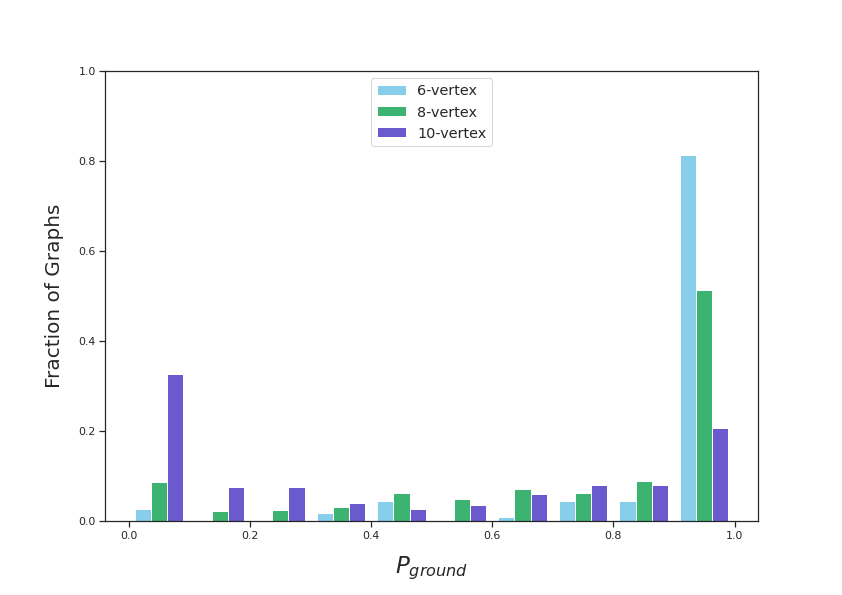}\label{fig:2b}}
    \caption{Histograms of (a) $C/C_{\text{max}}$ and (b) probability of overlap with ground state $P_{\text{ground}}$ after $s=4$ QITE steps with fixed Hamiltonian for all connected six- and eight-vertex graphs, and 200 randomly chosen ten-vertex graphs. The average values of $C/C_{\text{max}}$ are 0.99, 0.96, and 0.95, for six-, eight-, and ten-vertex graphs, respectively. The corresponding average values of $P_{\text{ground}}$ are 0.91, 0.66, and 0.48.}
    
    \label{fig:C4P4}
\end{figure}

\begin{figure}[ht!]
    \centering
        \subfigure[]{\includegraphics[scale=.3]{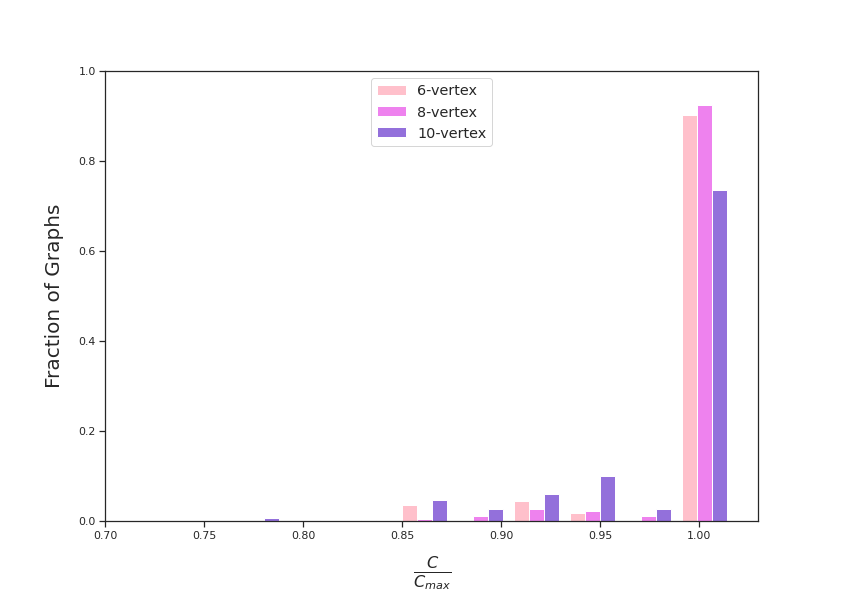}}
        \subfigure[]{\includegraphics[scale=.3]{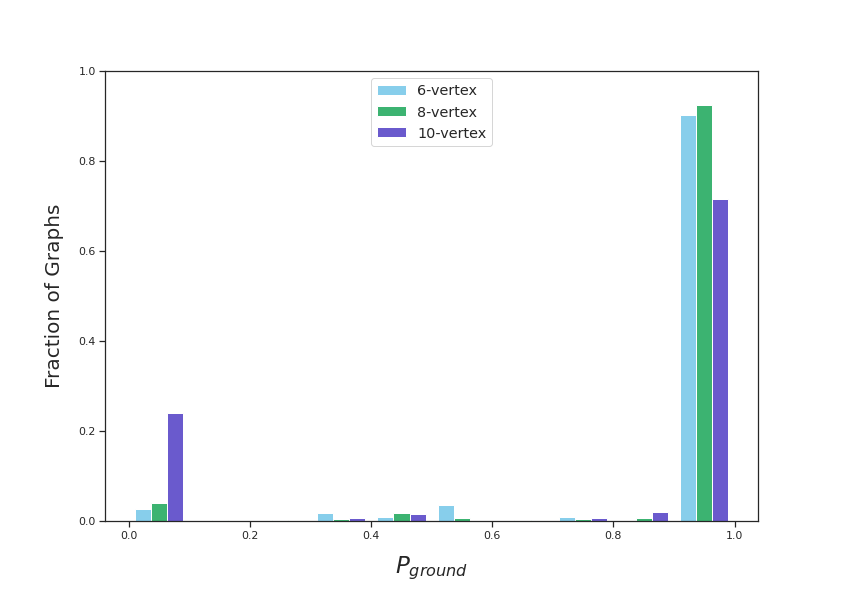}\label{fig:3b}}
    \caption{Histograms of (a) $C/C_{\text{max}}$ and (b) probability of overlap with ground state $P_{\text{ground}}$ after $s=10$ QITE steps with fixed Hamiltonian for all possible connected six- and eight-vertex graphs, and 200 randomly chosen ten-vertex graphs.   The average values of $C/C_{\text{max}}$ are 0.99, 0.98, and 0.97, for six-, eight-, and ten-vertex graphs, respectively. The corresponding average values of $P_{\text{ground}}$ are 0.94, 0.84, and 0.71. } \label{fig:C10P10}
\end{figure}

\begin{figure}[ht!]
    \centering
    \subfigure[]{\includegraphics[scale=.65]{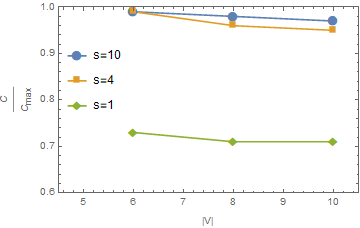}}
    \subfigure[]{\includegraphics[scale=.65]{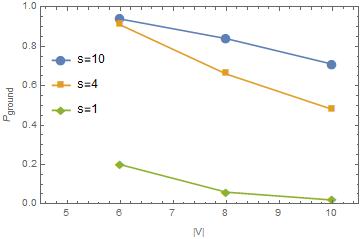}}
\caption{Average  (a) $C/C_{\text{max}}$ and (b) $P_{\text{ground}}$ for six-, eight-, and ten-vertex graphs after $s=1,4,10$ QITE steps.}
\label{fig:Cave}
\end{figure}

The performance of our algorithm improves dramatically after $s=4$ QITE steps, as shown in Figure \ref{fig:C4P4}. We obtained average values of $C/C_{\text{max}}$ 99\%, 97\%, and 94\%, and probabilities $P_{\text{ground}}$ 91\%, 74\%, and 45\%, for six-, eight-, and ten-vertex graphs, respectively. These results continued to improve with more QITE steps, although there is not much room for further improvement of $C/C_{\text{max}}$, as the values are very close to 100\%. At the tenth QITE step, the algorithm yields a final state which is very close to the one it converges to eventually. Figure \ref{fig:C10P10} shows that at $s=10$ QITE steps, the average values of $C/C_{\text{max}}$ are 99\%, 99\%, and 98\%, and average probabilities $P_{\text{ground}}$ are 94\%, 94\%, and 74\%, for six-, eight-, and ten-vertex graphs, respectively. A comparison of average $C/C_{\text{max}}$ and $P_{\text{ground}}$ values at $s=1,4,10$ QITE steps is summarized in Figure \ref{fig:Cave}. As expected, there is a degradation of performance as the number of vertices increases, but this appears to be a small effect.  In terms of the probability metric 
$P_{\text{ground}}$, we obtained a polarized histogram with a large accumulation at 100\% and a smaller one at 0\%. The 10\%, 20\%, 25\% of six-, eight-, ten-vertex graphs, respectively, with close to zero overlap with the ground state after $s=4$ QITE steps (Figure \ref{fig:2b}) remained there in subsequent QITE steps (see Figure \ref{fig:3b}). This is because the algorithm converges to an eigenstate of the Hamiltonian, which effectively terminates the algorithm as eigenstates are orthogonal to the ground state and do not update in the imaginary-time evolution \eqref{eq:9}.  For the majority of graphs, this is the ground state corresponding to the optimal solution $C_{\text{max}}$. However, for a small percentage of graphs, the algorithm converges to an excited energy level. This still yields a large value of the metric $C/C_{\text{max}}$, but the final state the algorithm converges to is an eigenstate of the Hamiltonian and therefore orthogonal to the ground state, which yields a value of the metric $P_{\text{ground}} = 0$.


One may try to gain insight into the cases in which QITE under-performs by analyzing the graphs for which the probability $P_{\text{ground}}$ is small. For example,
after $s=4$ QITE steps, about 20\% of the eight-vertex graphs 
have probability $P_{\text{ground}}$ below 0.2\%.  It is interesting to compare this result with the performance of classical algorithms, such as greedy algorithms \cite{10.1007/3-540-63774-5_137,articlegreedy,10.5555/1347082.1347102,7133122} or the Goemans-Williamson algorithm \cite{goemans1994879}. Classical (non-probabilistic) algorithms give $P_{\text{ground}} =0$ on graphs for which they fail to reach $C_{\textbf{max}}$. A comparison of the sets of graphs on which the greedy algorithms and the linear QITE algorithm, respectively, under-perform may elucidate the cause of such performance. Unfortunately, such a comparison does not appear to yield insights. For example, focusing on the Eulerian graphs, of which there are 184 out of a total of 11,117 connected eight-vertex graphs, after ten QITE steps, our method succeeded equally well on the Eulerian as well non-Eulerian graphs, unlike greedy algorithms.  Further simulations are needed to determine how the structure of a graph impacts the performance of our linear QITE algorithm. 

\begin{figure}[ht]
    \centering
        \subfigure[]{\includegraphics[scale=.2]{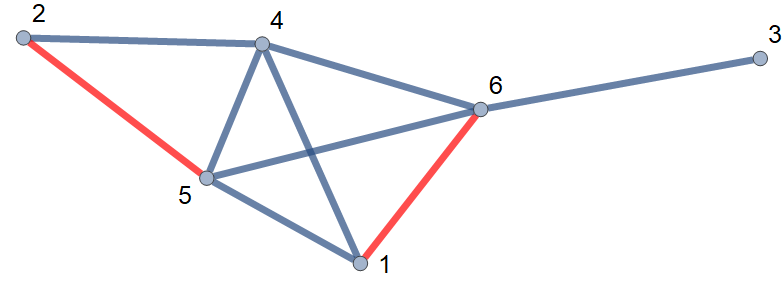}\label{fig6v}}
        \subfigure[]{\includegraphics[scale=.2]{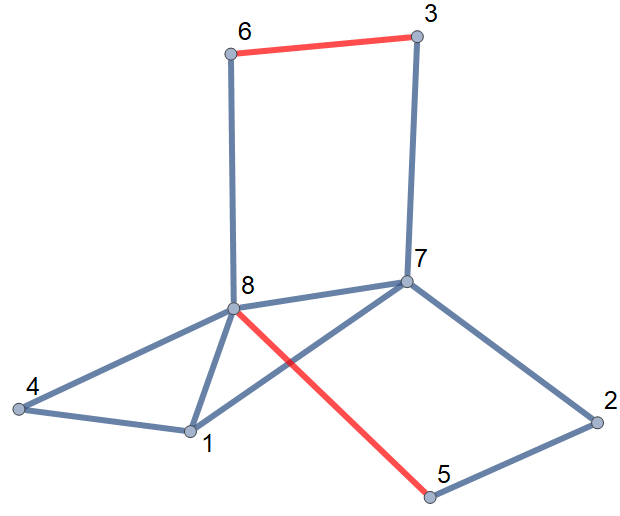}\label{fig8v}}
        \subfigure[]{\includegraphics[scale=.2]{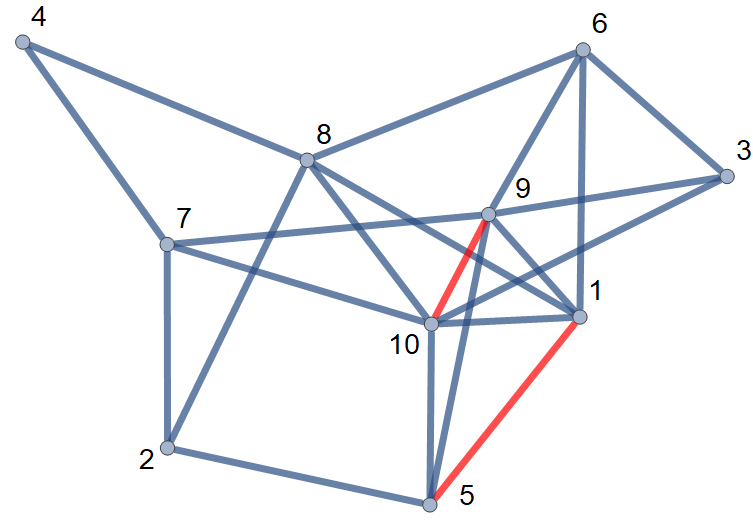}\label{fig10v}}
\caption{Example graphs on which our linear QITE algorithm under-performs: (a) 6-vertex, (b) 8-vertex, and (c) 10-vertex graphs.\label{fig:ExGph}}
\end{figure}

Next, we discuss examples of graphs with the worst performance of our linear QITE algorithm in terms of the $C/C_{\text{max}}$ metric with the probability metric $P_{\text{ground}} =0$. It should be pointed out that even in these worst-performing cases, we obtain values of the $C/C_{\text{max}}$ metric above 70\%. As we will show, it is possible to attain above $93\%$ performance in all cases in terms of both metrics $C/C_{\text{max}}$ and $P_{\text{ground}}$ with a slight modification of our algorithm.

\begin{figure}[ht]
    \centering
        \subfigure[]{\includegraphics[scale=.45]{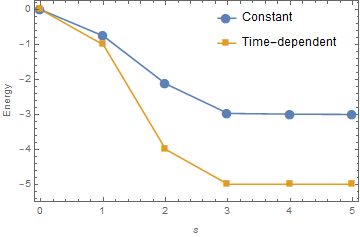}\label{fig:sixcomp}}
        \subfigure[]{\includegraphics[scale=.45]{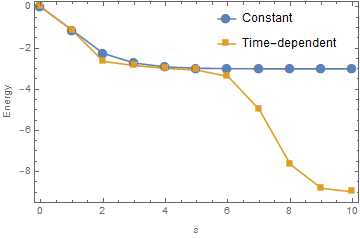}\label{fig:sixcompb}}
        \subfigure[]{\includegraphics[scale=.45]{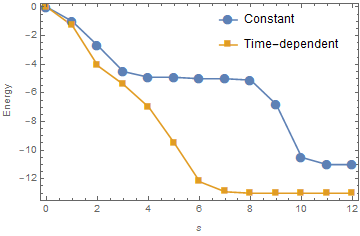}\label{fig:sixcompc}}
\caption{Convergence of our linear QITE algorithm for Hamiltonian with constant coefficients to energy levels $-3, -3, -11$, compared with imaginary-time-dependent Hamiltonian to ground state energy $-5,-9, -13$, for the six-, eight-, and ten-vertex graph, respectively, in Figure \ref{fig:ExGph}.}
\end{figure}

An example of worst-performing six-vertex graph is shown in Figure \ref{fig6v}. After applying $s=10$ QITE steps, we obtain $C=6$ whereas $C_{\text{max}}=7$. Thus, QITE has a performance of 86\% in terms of the $C/C_{\text{max}}$ metric. Moreover, the probability of overlap with the ground state is zero, and further application of our linear QITE algorithm will not improve its performance. This is because the algorithm converges to the state $\ket{0}^{\otimes 3} \otimes ( \cos\theta \ket{0} - \sin\theta \ket{1})^{\otimes 2}\otimes \ket{0}$, where $\tan\theta = \frac{1}{9}$, which is in the span of the first-excited states with energy $\mathcal{E}_{1} = -3$, whereas the ground state $\ket{110001}$ has energy $\mathcal{E}_{1} = -5$. 

An example of worst-performing eight-vertex graph is shown in Figure \ref{fig8v}. 
After applying $s=10$ QITE steps, we obtained $C = 7$, whereas $C_{\text{max}} =10$. Thus, $C/C_{\text{max}} = 0.7$. Once again, the probability of overlap with the ground state is zero, because our algorithm converges to the third excited state with energy $\mathcal{E}_{3} = -3$. More QITE steps will not move the state away from this eigenstate and towards the ground state of $\mathcal{H}$ (of eigenvalue $\mathcal{E}_0 = -9$). Ground states are $\ket{\bm{30}} = \ket{00011110}$ and $\ket{\bm{225}} = \ket{11100001}$ (written in binary notation), where the latter is obtained from the former by flipping all qubits. The state we end up with after $s=10$ QITE steps is $\ket{0001}\otimes (\cos\theta\ket{0}+\sin\theta\ket{1})^{\otimes 2} \otimes \ket{11}$, where $\tan\theta = 0.5$. It is a linear combination of the states $\ket{\bm{19}}, \ket{\bm{23}}, \ket{\bm{27}}, \ket{\bm{31}}$, all of which are eigenstates of the Hamiltonian with corresponding eigenvalue $\mathcal{E}_{3} = -3$. 

An example of worst-performing ten-vertex graph is shown in Figure \ref{fig10v}. 
We obtained convergence to $C = 16$, whereas $C_{\text{max}} =17$. Thus, $C/C_{\text{max}} = 0.94$ whereas the probability of overlap with the ground state is zero, because our algorithm converged to the first excited state $\ket{\bm{188}} = \ket{0010111100}$ with energy $\mathcal{E}_{1} = -11$ which is orthogonal to the ground states $\ket{\bm{339}} = \ket{0101010011}$ and $\ket{\bm{684}} = \ket{1010101100}$ with energy $\mathcal{E}_0 = -13$. 

Interestingly, the performance can be improved further by considering an imaginary-time-dependent (ITD) Hamiltonian (Eq.\ \eqref{eq:Hbeta}). To this end, we start the time evolution along a different path in terms of updates with respect to the imaginary-time-dependent $\mathcal{H}_{G_j} [s]$, thereby evading the standard path which results in convergence to an excited state. Let $\mathcal{H}_G$ be the Hamiltonian corresponding to the graph $G = (V,E)$ of interest, and $\mathcal{H}_{G'}$ the Hamiltonian for a subgraph $G' = (V',E')$ of $G$. Initially, we excise $G'$ and gradually turn it on to form the graph $G$ we are interested in. This leads us to consider the Hamiltonian with ITD edges,
\beq \mathcal{H} [s] = \mathcal{H}_G - f(s) \mathcal{H}_{G'} \eeq
where $f(s)$ interpolates between $1$ for $s=1$ and $0$ for large $s$ ($s \ge s_0$ for a given $s_0$), interpolating between the graph $G$ and its subgraph in which $G'$ has been excised. In terms of the weights $h_{ij}$ in Eq.\ \eqref{eq:Hbeta}, we have constant weights $h_{ij} =1$, for edges not in $G'$ ($(ij) \notin E'$), and $h_{ij} = 1- f(s)$, for $(ij) \in E'$.

For the six-vertex graph in Figure \ref{fig6v}, by excising two edges ($E' = \{ (16), (25) \}$), and switching them back on in three QITE steps ($f(1)=1$, $f(2) = 0.5$, and $f(s) = 0$ for $s>2$), we obtained convergence to the ground state, and therefore performance of 100\% in both metrics $C/C_{\text{max}}$ and $P_{\text{ground}}$. A comparison between the Hamiltonian with constant coefficients and the interpolating Hamiltonian is shown in Figure \ref{fig:sixcomp}.
For the subgraph $G'$ to be excised, we found 11 different possibilities consisting of pairs of edges out of $\binom{9}{2}=36$ possible combinations for which our algorithm successfully converged to the ground-state.  

Similarly, for the eight-vertex graph in Figure \ref{fig8v}, we obtained convergence to the ground state by excising $G'$ consisting of the pair of edges $E' = \{ (36), (58) \}$. It turned out that out of all $\binom{11}{2} = 55$ combinations, 17 yielded convergence to the ground state. For the ten-vertex graph in Figure \ref{fig10v}, convergence to the ground state was obtained for 40 different $G'$ subgraphs consisting of pairs of edges out of $\binom{18}{2} = 153$ possible combinations. A comparison between the Hamiltonian with constant coefficients and the interpolating Hamiltonian is shown in Figures \ref{fig:sixcompb} and \ref{fig:sixcompc} for the eight- and ten-vertex graphs, respectively.


Motivated by the success of the modified linear QITE algorithm with ITD Hamiltonian on the worst-performing graphs in Figure \ref{fig:ExGph}, we applied the strategy of gradually switching on a pair of edges to the remaining graphs on which our linear QITE algorithm under-performed, i.e, it was not 100\% successful and did not converge to the ground state that would yield the optimal MaxCut solution.
Out of all 112 connected six-vertex graphs, our linear QITE algorithm using a Hamiltonian with constant weights led to convergence to the ground state for 101 of them. For the remaining 11 graphs, we obtained convergence to the first excited state. Using an ITD Hamiltonian with an appropriate choice of a pair of ITD weighted edges, we obtained convergence to the ground state for all remaining 11 graphs.

\begin{figure}[ht!]
    \centering
        \subfigure[]{\includegraphics[scale=.3]{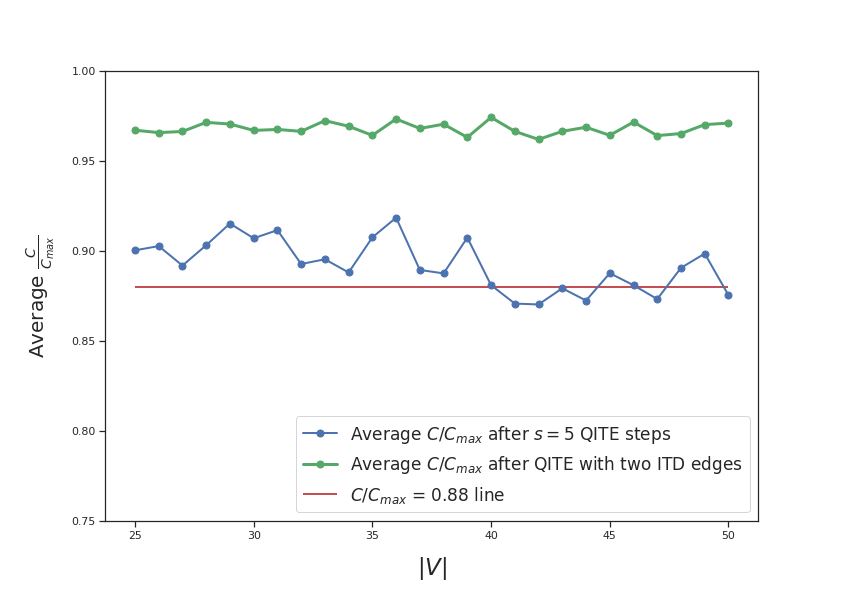}\label{fig:AveC}}
        \subfigure[]{\includegraphics[scale=.3]{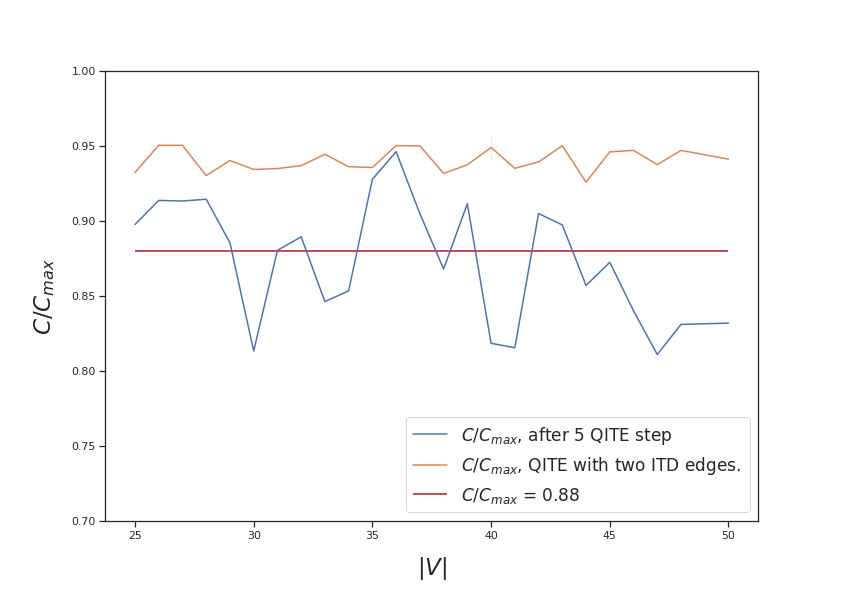}\label{fig:worstperf}}
    \caption{(a) Average ${C}/{C_{\textrm{max}}}$ \textit{vs.}\ number of vertices. The blue line is the result of applying $s=5$ QITE steps, and the green line is the result of applying QITE with two ITD edges. (b) $C/C_{\textrm{max}}$ \textit{vs.}\ number of vertices for the worst performing graphs computed with five-step linear QITE (blue line) and modified QITE with ITD edges (orange line). For reference to classical algorithms, the 88\% worst-performance level of the Goemans-Williamson algorithm is indicated with a red line.}
    
    
    \label{}
\end{figure}

\begin{figure}[ht!]
    \centering
    \includegraphics[scale=.3]{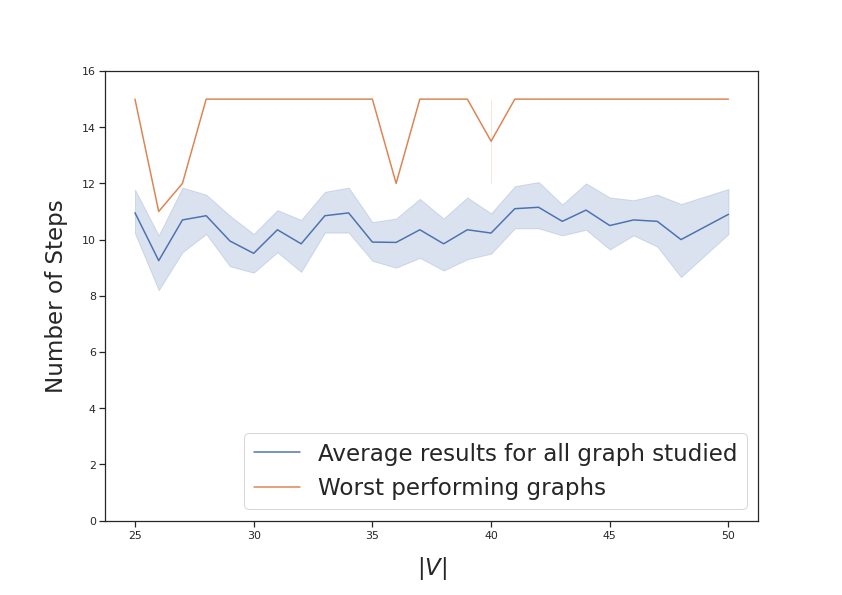}
    \caption{Average number of steps to reach $C/C_{\textrm{max}}$ of at least 93\% \textit{vs.} number of vertices.}
    \label{fig:AveStep}
\end{figure}

Similar results were obtained for eight-vertex graphs. Our linear QITE algorithm with constant weights led to convergence to the ground state for 8,995 graphs out of all 11,117 connected eight-vertex graphs. For the remaining 2,122 graphs, we obtained convergence to an excited state, mostly the first excited state, which explains the near-perfect performance in terms of the $C/C_{\text{max}}$ metric. By switching on appropriately chosen pairs of edges, our modified linear QITE algorithm led to convergence to the ground state in all remaining 2,122 eight-vertex graphs. A pair of \ra{ITD} edges also sufficed for ten-vertex graphs that we considered. Out of a randomly chosen sample of 200 graphs, 134 converged to the ground state with our algorithm using constant weights. The remaining 76 ten-vertex graphs also converged to the ground state after excising and gradually switching on a pair of edges that were appropriately chosen in each case.  

\begin{figure}[ht!]
    \centering
        \subfigure[]{\includegraphics[scale=.3]{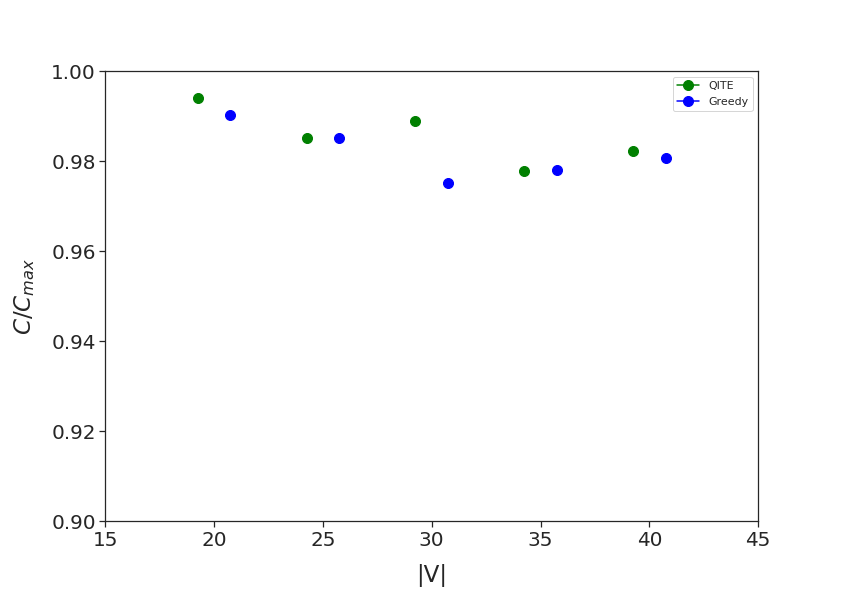}\label{fig:GreedyComp}}
        \subfigure[]{\includegraphics[scale=.255]{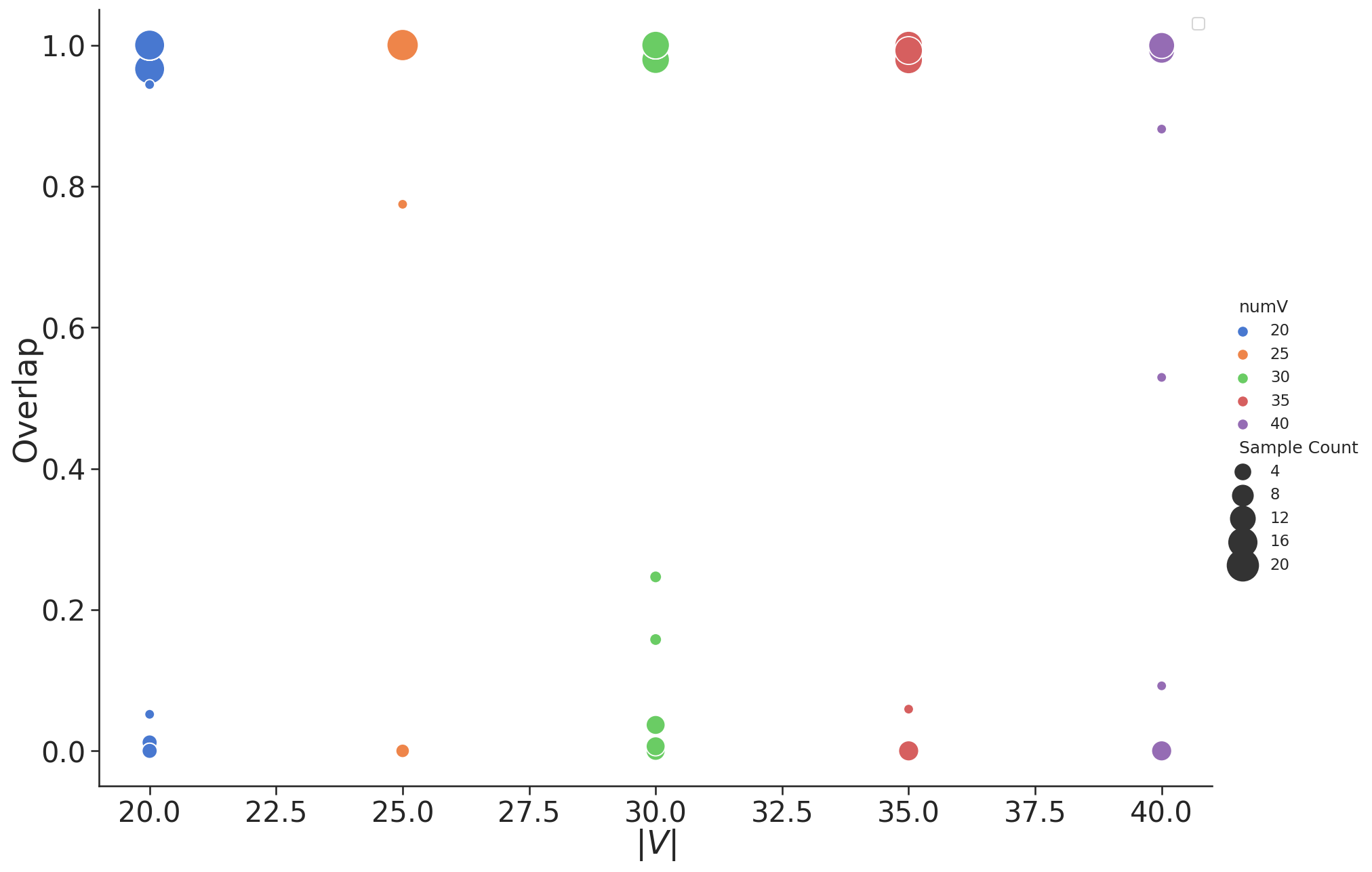}\label{fig:OvlpDist}}
    \caption{(a) Average ${C}/{C_{\textrm{max}}}$ comparison between QITE and a classical greedy algorithm. (b) Overlap with ground state for a random sample of graphs with number of vertices $|V| = \{20, 25, 30, 35, 40\}$. }
    
    \label{fig:ovlp}
\end{figure}

Even though a large number of pairs of edges leads to convergence to the ground state, there appears to be no way of determining them from the graph. However, going through all possible combinations only adds a polynomial overhead ($\mathcal{O} (|V|^2)$) to the calculation. It would be interesting to determine how the number of edges that need to be assigned weights that vary with imaginary time changes as larger graphs are analyzed. Our results indicate that our linear QITE method offers an efficient solution to the MaxCut problem with complexity of the algorithm growing polynomially with the number of vertices $|V|$ of the graph.

The performance of our linear QITE algorithm remains robust on larger graphs. We applied the modified linear QITE algorithm with ITD edges to over 1,000 randomly chosen graphs with up to 50 vertices. The results are summarized in Figure \ref{fig:AveC}. The blue line shows the performance of five steps of linear QITE whereas the green line shows the performance of the modified algorithm with two ITD edges showing significant performance improvement. The lowest $C/C_{\textrm{max}}$ was found to be above 93\%. This is above the performance of the Goemans-Williamson algorithm \cite{goemans1994879}, which is the best known classical algorithm, at 88\% approximation ratio for MaxCut. 

Accumulated results for the graphs we analyzed with between 25 and 50 vertices are plotted in Figure \ref{fig:AveStep} (blue line). It is important to note that the number of QITE steps needed to reach $C/C_{\textrm{max}}$ above 93\% is uniform, varying between 9 and 11, and does not grow with the number of vertices of the graphs. A performance comparison after applying $s=5$ QITE steps to the worst performing graphs with the modified QITE method employing two ITD edges is shown in Figure \ref{fig:worstperf}.




All the graphs we analyzed above have been randomly chosen with edge probability between 9\% and 99\%. Next, we focus on less dense graph with edge probability randomly chosen between 3.5\% and 4.5\% per vertex which produces four edges per vertex on average. Figure \ref{fig:GreedyComp} shows a comparison between QITE and a classical greedy algorithm \cite{7133122}. Each green data point represents the average $C/C_{\text{max}}$ over 25 different graphs calculated using QITE on randomly chosen graphs with the same number of vertices $|V|$ for $|V| \in \{ 20, 25, 30, 35, 40 \}$. We applied the modified QITE algorithm with ITD edges making 100 random selections of pairs of such edges from a total of $\mathcal{O} (|V|^2)$ pairs and choosing the best result. The blue data points represent average $C/C_{\text{max}}$ values of the same graphs obtained with a greedy algorithm. As we can see, the QITE performance is comparable to the performance of the classical greedy algorithm. Unlike the deterministic greedy algorithm, QITE is a probabilistic algorithm that yields a superposition state that we can take advantage of. QITE may converge to a superposition state that has a finite overlap with the ground state. In this case, one can obtain $C_{\textrm{max}}$ even when the average $C/C_{\text{max}}$ is less than 100\%. Figure \ref{fig:OvlpDist} shows the overlap with the ground state for the same graphs. Only four to seven among 25 graphs of a given number of vertices has zero overlap with the ground state. The rest of the graphs have non-vanishing overlap with most between 90\% and 100\% percent.


\section{Conclusion}\label{conclusion}

The ability of quantum algorithms to outperform their classical counterparts is primarily due to their ability to to tap the resource of quantum entanglement. In the case of combinatorial optimization, this is seen in QAOA \cite{farhi2014quantum} which introduces entanglement to solve a problem, such as MaxCut, in which both the initial state and desired final state are separable. However, despite considerable effort, quantum advantage is yet to be proved or demonstrated experimentally. QITE offers a different approach to combinatorial optimization by effectively cooling the system down to its ground state and obtaining its minimum energy that corresponds to the optimal solution of the corresponding combinatorial optimization problem \cite{motta2020determining}. In general, QITE also introduces entanglement, but quantum advantage is yet to manifest.

In an effort to investigate the impact of entanglement on combinatorial optimization problems, we applied a version of QITE to the MaxCut problem that used a linear \emph{Ansatz} for unitary updates and a separable initial state. Thus, we introduced no entanglement in the quantum algorithm which could be efficiently simulated by a classical computer. Remarkably, even though the linear updates introduced approximations at each step, for all graphs we analyzed our linear QITE algorithm succeeded in converging to the ground state, thus resulting in a  performance exceeding 93\% in terms of the metric $C/C_{\text{max}}$. Importantly, the number of steps needed to reach that performance remained fairly uniform over the graphs we analyzed and did not grow with the number of vertices.
Although further work on higher-order graphs is needed, our results indicate that our linear QITE method is efficient with the number of steps growing modestly with the number of vertices $|V|$ of the graph.
In detail, we analyzed graphs with up to fifty vertices. After ten QITE steps using a Hamiltonian with constant weights, all graphs almost converged to an eigenstate of the Hamiltonian. A high percentage of them converged to the ground state. The remaining graphs converged to an excited state which, however, still led to very high performance in terms of the average $C/C_{\text{max}}$ metric. The average $C/C_{\text{max}}$ was found to be 89\%. Furthermore, we were able to improve this performance and obtain convergence to the ground state, leading to  average $C/C_{\text{max}} \approx 97\%$ and worst performance of 93\%, by considering an imaginary-time-dependent Hamiltonian resulting from excising a pair of edges appropriately chosen and switching them back on gradually within a few QITE steps. For each graph, we found several pairs of edges that led to the ground state. However, there appears to be no guidance on how to choose these edges in a given graph. Nevertheless, checking all possible pairs of vertices only introduces a polynomial overhead ($\mathcal{O} (|V|^2)$) in the calculation.

To observe quantum advantage in the solution of the MaxCut problem, it appears to be necessary to consider graphs that are much larger than  fifty-vertex graphs. This may take us outside the realm of NISQ devices as the depth of the quantum circuits will introduce a prohibitive amount of quantum errors. On the other hand, our work can be extended to larger graphs on which we can perform numerical calculations and investigate how complexity depends on the number of vertices and the role of entanglement. Entanglement can be introduced by adding higher-order terms to the unitary updates at each QITE step. It would be interesting to study what type of graphs with a large number of vertices can be accessed with NISQ devices at performance levels that exceed other quantum algorithms. Work in this direction is in progress.

\acknowledgments

This work was supported by the DARPA ONISQ program under award W911NF-20-2-0051. J.\ Ostrowski acknowledges the Air Force Office of Scientific Research award, AF-FA9550-19-1-0147. G.\  Siopsis  acknowledges the Army Research Office award W911NF-19-1-0397 and the National Science Foundation award DGE-2152168. J.\ Ostrowski and G.\ Siopsis acknowledge the National Science Foundation award OMA-1937008.

This manuscript has been authored by UT-Battelle, LLC under Contract No.\ DE-AC05-00OR22725 with the U.S.\ Department of Energy. The United States Government retains and the publisher, by accepting the article for publication, acknowledges that the United States Government retains a non-exclusive, paid-up, irrevocable, world-wide license to publish or reproduce the published form of this manuscript, or allow others to do so, for United States Government purposes. The Department of Energy will provide public access to these results of federally sponsored research in accordance with the DOE Public Access Plan. (http://energy.gov/downloads/doe-public-access-plan).

\end{document}